# A Spatial-Physics Informed Model for 3D Spiral Sample Scanned by SQUID Microscopy

J. Senthilnath[1*], Jayasanker Jayabalan[2*], Zhuoyi Lin[1], Aye Phyu Phyu Aung[1], Chen Hao[1], Kaixin Xu[1], Yeow Kheng Lim[2], F. C. Wellstood[3,4*]

[1]Institute for Infocomm Research (I²R), Agency for Science, Technology and Research (A*STAR), Singapore
[2]Department of Electrical Engineering, National University of Singapore, Singapore
[3]Neocera Magma, Beltsville, MD, USA
[4]Center for Nanophysics and Advanced Materials, Department of Physics, University of Maryland, College Park, MD, USA
*Corresponding authors: J_Senthilnath@i2r.a-star.edu.sg, jayajaya@nus.edu.sg, well@umd.edu

*Abstract*—The development of advanced packaging is essential in the semiconductor manufacturing industry. However, non-destructive testing (NDT) of advanced packaging becomes increasingly challenging due to the depth and complexity of the layers involved. In such a scenario, Magnetic field imaging (MFI) enables the imaging of magnetic fields generated by currents. For MFI to be effective in NDT, the magnetic fields must be converted into current density. This conversion has typically relied solely on a Fast Fourier Transform (FFT) for magnetic field inversion; however, the existing approach does not consider eddy current effects or image misalignment in the test setup. In this paper, we present a spatial-physics informed model (SPIM) designed for a 3D spiral sample scanned using Superconducting QUantum Interference Device (SQUID) microscopy. The SPIM encompasses three key components: i) magnetic image enhancement by aligning all the "sharp" wire field signals to mitigate the eddy current effect using both in-phase (I-channel) and quadrature-phase (Q-channel) images; (ii) magnetic image alignment that addresses skew effects caused by any misalignment of the scanning SQUID microscope relative to the wire segments; and (iii) an inversion method for converting magnetic fields to magnetic currents by integrating the Biot-Savart Law with FFT. The results show that the SPIM improves I-channel sharpness by 0.3% and reduces Q-channel sharpness by 25%. Also, we were able to remove rotational and skew misalignments of $0.3^0$ in a real image. Overall, SPIM highlights the potential of combining spatial analysis with physics-driven models in practical applications.

*Keywords—Magnetic field Imaging, physics informed model, eddy current, affine transform, fast fourier transform*

## I. Introduction

In the semiconductor packaging industry, various microscopy scanning techniques are employed for failure analysis (FA). These include physical failure analysis methods such as Scanning Electron Microscopy (SEM) and Transmission Electron Microscopy (TEM) [1], as well as non-destructive testing methods like Lock-in Thermography (LiT) and X-ray imaging [2]. Integrated circuit (IC) manufacturers commonly use vision-based semiconductor microscopies, including optical microscopy, SEM, and X-ray imaging, to monitor wire segments. However, there is a growing demand for non-visual inspection methods, as most FA technologies currently lack the capability for non-contact imaging of buried currents. To address these challenges, magnetic field imaging (MFI) [3] is being explored, although it is still in the research and development stage within the semiconductor field.

MFI-derived magnetic images can provide information about the lateral and vertical resolution, as well as the coordinates and direction of current flow. This method is primarily used to identify buried electrical defects at both the device and packaging levels. MFI allows for the imaging of magnetic fields generated by currents, even through relatively thick layers of insulation and metal. Three popular MFI techniques are Giant MagnetoResistance (GMR), Quantum Diamond Microscope (QDM), and Superconducting QUantum Interference Device (SQUID). Among these, SQUID can be designed to measure magnetic fields as small as 1 femtotesla (1 fT or $10^{-15}$ T), which surpasses the sensitivity of QDM (approximately 10 pT/√Hz) and GMR sensors (approximately 1 nT/√Hz) for detecting weak signals [4] [5] [6]. For magnetic microscopy, it is essential to achieve both very good field resolution and a compact sensor size. The SQUID provides a good balance in this regard. It can detect currents as low as 100 nA at distances greater than 500 μm, allowing non-contact imaging of buried currents. Its high sensitivity also allows SQUID to image leakage paths under low test currents, reducing the risk of damaging or "healing" sensitive defects [4]. Operating at cryogenic temperatures, SQUIDs are not affected by thermal drift, whereas QDM requires temperature control to prevent Joule heating effects. SQUIDs are also less influenced by packaging materials, while QDM suffers from strong background gradients caused by magnetized C4 bumps [6]. Furthermore, SQUIDs produce noise as low as 8 pT/√Hz, whereas GMR sensors exhibit noise of 1 nT/√Hz [4]. SQUID produces higher-quality magnetic field images than GMR, making it well-suited for current path reconstruction through magnetic inversion techniques [5]. Additionally, SQUID imaging benefits from mature current reconstruction techniques, whereas QDM's inverse problem remains more complex [6]. This study focuses on SQUID-based microscopy scanned image.

SQUID microscopy allows for imaging magnetic fields generated by currents flowing through ICs. Nonetheless, there are several challenges associated with this method, such as eddy current effects that may arise from sample placement, the operating frequency, and the magnetic fields produced by good conducting materials [7]. Additional processing steps are necessary before converting magnetic field images into current density images. These include removing eddy current effects and correcting circuit misalignment during scanning with the SQUID. To address these challenges, we propose a spatial-physics informed model (SPIM) to improve spatial resolution [8]. To the best of our knowledge, there is currently no published literature that integrates magnetic image enhancement, magnetic image alignment, and magnetic image conversion.

In this paper, we present a spatial-physics informed model (SPIM) that operates in three stages: i) Enhancing the in-phase (I-channel) image to be significantly sharper than the quadrature-phase (Q-channel) image, ii) Applying an affine transformation to correct for rotation and skew effects, and iii) Combining the Biot-Savart Law and Fast Fourier Transform (FFT) to convert the magnetic field data into current density. We applied the SPIM method to our experimental MFI data, which involved scanning a 3D spiral sample using SQUID microscopy. The SPIM method demonstrates superior enhancement and effectively converts magnetic field data into magnetic current, making it particularly suitable for semiconductor failure analysis applications.

## II. SPATIAL-PHYSICS INFORMED MODEL

The proposed spatial-physics informed model (SPIM) method operates in three stages for magnetic image conversion of current-carrying circuits into the current density. Figure 1 shows an overview of the SPIM.

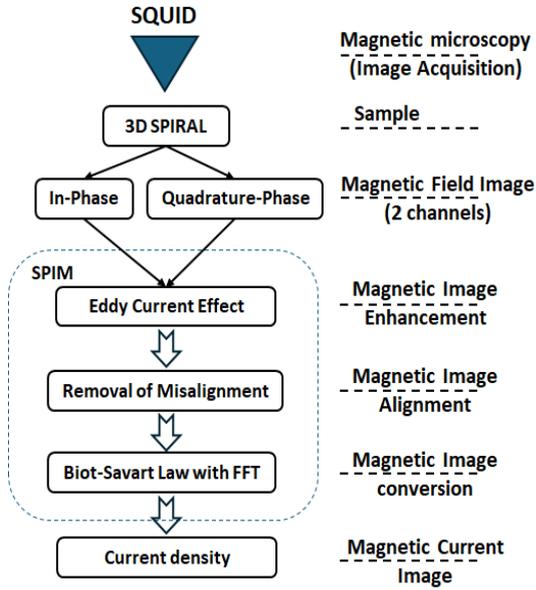

Fig. 1. An overview of the SPIM

The signal recorded in a SQUID data file consists of the in-phase ($V_I$, I-channel) and quadrature-phase ($V_Q$, Q-channel) output voltages ($V_{out}$) from a lock-in detector. This detector is connected to the output ($V_f$) of a SQUID Flux-Locked Loop (FLL) measuring system, as shown in Figure 2 [8, 9]. $V_{out}$ is proportional to the amplitude of the magnetic field at the SQUID, resulting in two representations of the sample: the I image and the Q image.

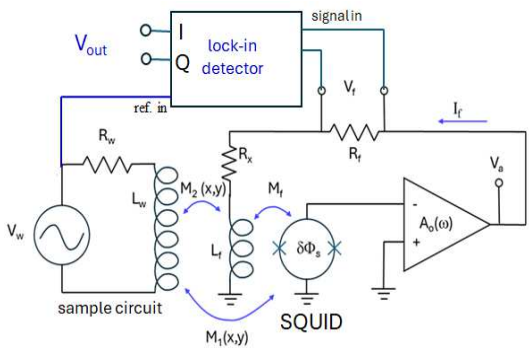

Fig. 2. Schematic showing setup for SQUID imaging

### A. Magnetic Image Enhancement

The magnetic image enhancement is performed by aligning all the "sharp" wire field signals to mitigate the eddy current effect using both in-phase (I-channel) and quadrature-phase (Q-channel) images. Initially, the I-channel and Q-channel images are converted from output voltage to magnetic field in Tesla using the conversion factor $\frac{dB_z}{dV}$ which is given by,

$$B_z = \frac{dB_z}{dV} * V_{out} = \frac{V_{out}}{A_s \frac{dV}{d\Phi}} \approx (3.632\ \mu T/V) * V_{out} \quad (1)$$

where $\frac{dV}{d\Phi} = \frac{R_f}{M_f} \approx 0.556\ \frac{V}{\Phi_0} \approx 2.687\ \text{x}10^{14}\ \text{V/(Tm}^2\text{)}$, $R_f$ is a feedback resistance in Flux-Locked Loop (FLL) $\approx 5\ k\Omega$, $M_f$ is a feedback mutual inductance in FLL $\approx 19\ pH$, $\Phi_o = \frac{h}{2e} =$ flux quantum $= 2.07\text{x}10^{-15}\ \text{Tm}^2$, SQUID pickup area $A_s = d^2 \approx (32\ \mu m)^2$, and side-length of SQUID pick-up loop: $d \approx 32\ \mu m$.

Ideally, before making an image, the user adjusts the reference phase of the lock-in so that the I-channel contains all the signal, and the Q-channel is just noise (phase is set to zero). The I-channel can then be used as "the magnetic image", and the Q-channel can be ignored. This is the case of zero phase. However, images of real samples often show signals in both channels and further processing may be needed before a proper magnetic image can be fed to the magnetic image alignment and magnetic image conversion.

<u>Simple case of constant non-zero phase</u>: this can happen if the reference phase of the lock-in was not adjusted to move the signal into just the I-channel. In this case, we can use below equation to reset the phase so that all the magnetic signal ends up in a new "rotated" I'-channel. Rotate the lock-in phase ϕ to put all the "sharp" wire field signal into I' image and just noise or "blurry" eddy currents in Q' using

$$I' = I \cos(\phi) + Q \sin(\phi)$$

$$Q' = -I \sin(\phi) + Q \cos(\phi) \quad (2)$$

Note, if "Eddy-currents" are present, the local phase of the signal ϕ=arctan(Q/I) can vary systematically with position.

### B. Magnetic Image Alignment

Misalignment in the test setup, such as rotation or skew effects, can cause shifts in raster-scanned images. An affine transformation, which overcomes rotation or skew effects, is used here to correct these geometric distortions with the necessary parameters to properly align the images [10, 11].

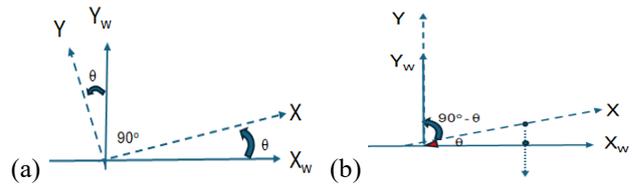

Fig. 3. SQUID scanned raw magnetic image with (a) rotation effect, (b) skew effect.

The image transformation (T) of I' by applying rotation or skew (A) using,

$$T = A \times I' \quad (3)$$

Imaging using rotated scanning axes: Let (X, Y) be the coordinates recorded for the pixels in I' image taken while scanning using axes that are rotated with respect to the wires (Figure 3(a)). Then the position in real space T with axes along and perpendicular to the direction of the wires is ($X_w$, $Y_w$) where:

$$X_W = X \cos(\theta) - Y \sin(\theta)$$
$$Y_W = X \sin(\theta) + Y \cos(\theta) \quad (4)$$

Imaging using skewed axes: Let (X, Y) be the coordinates obtained from I' image taken while scanning along skewed axes (Figure 3(b)). Then the position in real space T with axes along X and Y direction of the perpendicular wires is ($X_w$, $Y_w$) where:

$$X_w = X$$
$$Y_w = Y + X \tan(\theta) \quad (5)$$

For the magnetic image conversion, we need to find the field at the real space position ($X_w$, $Y_w$) of each pixel.

### C. Magnetic Image Conversion

The Biot-Savart law takes you from current density **J** to magnetic field **B**. Suppose there is a 2D current distribution **J**(x,y) in the xy plane that is a distance $z$ away then:

$$B_z(x,y,z) \approx \frac{\mu_0 d}{4\pi} \int_{-\infty}^{\infty} \int_{-\infty}^{\infty} \frac{J_x(x',y') \cdot (y-y') - J_y(x',y') \cdot (x-x')}{[(x-x')^2 + (y-y')^2 + z^2]^{3/2}} dx' dy' \quad (6)$$

Take the 2D Fourier transform of the Biot-Savart Law for $B_z(x,y)$ to get [3],

$$b_z(k_x, k_y, z) = i \frac{\mu_0 d}{2} \frac{e^{-kz}}{k}[k_y \cdot j_x(k_x, k_y) - k_x \cdot j_y(k_x, k_y)] \quad (7)$$

For a 2D current distribution $\bar{J}(x,y)$ in the xy plane, we have $\nabla° \bar{J} = 0$. Taking the Fourier transform gives:

$$-ik_x j_x(k_x, k_y) - ik_y j_y(k_x, k_y) = 0.$$

Combining Equations (6) and (7), in general [3],

$$j_x(k_x, k_y) = -\frac{2i}{\mu_0 d} e^{kz} \frac{k_y}{k} b_z(k_x, k_y, z) \quad (8)$$

$$j_y(k_x, k_y) = +\frac{2i}{\mu_0 d} e^{kz} \frac{k_x}{k} b_z(k_x, k_y, z) \quad (9)$$

where $k = \sqrt{k_x^2 + k_y^2}$

Take inverse fourier transform to get real space current density images

$$J_x(x,y) = -\frac{1}{(2\pi)^2} \iint \frac{2i}{\mu_0 d} \frac{k_y}{k} e^{kz} b_z(k_x, k_y, z) e^{-i(k_x x + k_y y)} \cdot f_x(k_x, k_y) dk_x dk_y \quad (10)$$

$$J_y(x,y) = \frac{1}{(2\pi)^2} \iint \frac{2i}{\mu_0 d} \frac{k_x}{k} e^{kz} b_z(k_x, k_y, z) e^{-i(k_x x + k_y y)} \cdot f_y(k_x, k_y) dk_x dk_y \quad (11)$$

where $k_x$, $k_y$, and $k$ are real, dimensioned, physical quantities with units of $m^{-1}$, while $k*z$ is dimensionless.

### III. EXPERIMENTAL RESULTS

In this section, we discuss the experimentally obtained real MFI data and magnetic image analysis using SPIM.

#### A. Experimental Set-Up of the 3D spiral sample

Figure 4 provides the 3D visualization of our spiral sample that shows an ordered connected wire path or current trace ($P_0$, $P_1$,..., $P_{12}$). Figure 5 illustrates the top layer of the spiral sample with the current input and output. The magnetic image consists of two channels: in-phase (I-channel) and quadrature-phase (Q-channel) images. The sample scan time is approximately 15 minutes. Figure 6 presents the magnetic image (I-channel) generated by scanning the 3D spiral sample with SQUID microscopy, which is overlaid on the spiral sample image. It is important to note that some wires may overlap. The image dimensions are 254 x 1166 pixels, with a total of 296,164 data points, and each pixel measures 30 μm. The magnetic image exhibits a significant eddy current signal along with position noise. Note, wires in the $z$ direction produce no $z$-component of the magnetic field.

TABLE I. 3D SPIRAL SAMPLE DESCRIPTION

| Sample description | Value |
|---|---|
| Number of layers | 2 |
| Current (I) | 1000 μA |
| Frequency | 5000 Hz |
| SQUID noise | 0.1 nT |
| All lateral wires have with ℓ/z | >> 1 |
| Wire width* | 200 μm |
| Number of paths | ~9 segments |
| Large separation | ~200 μm |
| SQUID to surface of the sample | ~120 μm |

*manufacturing tolerance of ±10% of the given wire width

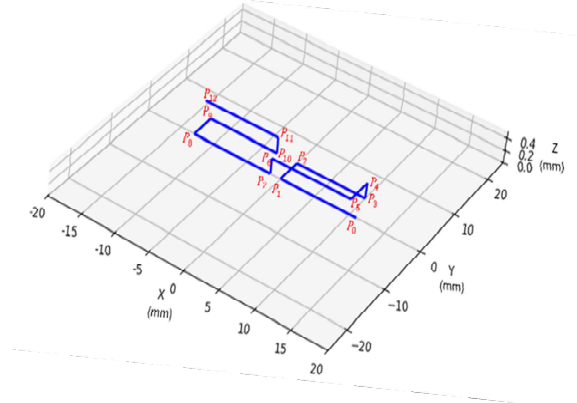

Fig. 4. 3D visualization of the connected wire path.

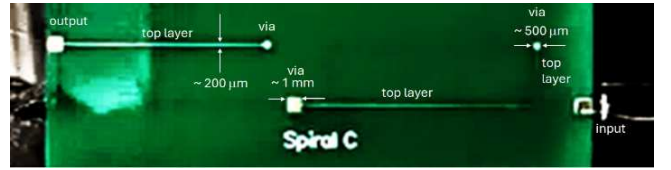

Fig. 5. 3D spiral sample image.

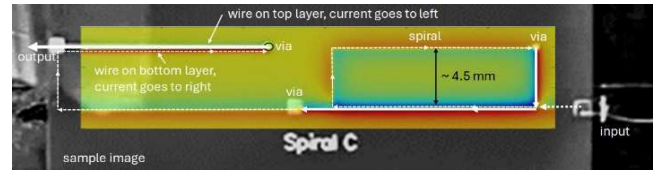

Fig. 6. SQUID scanned magnetic image overlaid on spiral sample image.

#### B. Magnetic Image Analysis

Figure 7(a) shows the SQUID-scanned I-channel and Figure 7(b) shows the Q-channel images in Volts for the 3D spiral sample. Figure 8(a) shows the converted raw, I-channel and Figure 8(b) shows the Q-channel images, where the Volts have been transformed into magnetic field values expressed in Tesla, following Eq. (1).

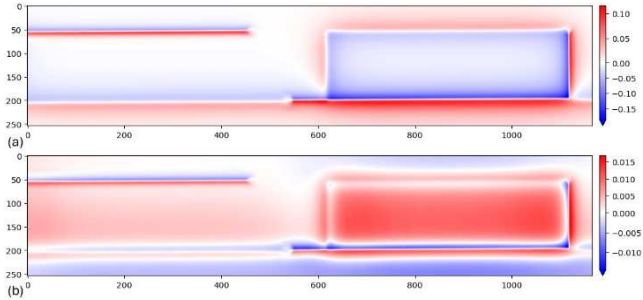

Fig. 7. SQUID scanned raw magnetic image in Volts with uncorrected phase (a) I-channel image in V, (b) Q-channel image in V.

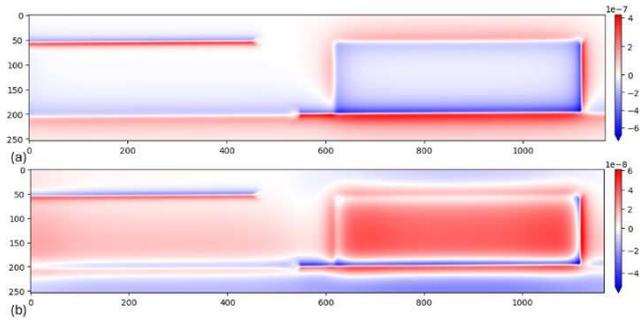

Fig. 8. Converted from Volts to Tesla (a) I-channel image in T, (b) Q-channel image in T.

Total Variation (TV) score: To evaluate the effect of phase-value, we use the discrete form of TV score [12], defined as:
$$TV(u) = \sum_{i,j} \sqrt{(u[i+1,j] - u[i,j])^2 + (u[i,j+1] - u[i,j])^2} \quad (12)$$
where $u(i,j)$ is the output image from the eddy current analysis.

Figure 9 presents an analysis of tuning phase values aimed at maximizing the TV score using the I′ image. In Figure 9(a), the TV scores for our magnetic field I′ image are relatively low, approximately in the range of $10^{-3}$. This indicates that our eddy current analysis of the I′ image is significantly sharper than that of the Q′ image, which minimizes sharp features. In Figure 9(b), we narrow down the analysis by selecting the top two high TV scores (5 and 6, as shown in Figure 9(a)) with a step of 0.2, moving from phase value 5 to 6. After conducting a more detailed analysis of the phase values, we identify 5.8 as having the highest TV score. This selection ensures that the sharp features in the I′ image are preserved while minimizing sharp features in the Q′ image (notably, the $B_z$ scale of the Q′ image is 10x smaller than that of the I′ image).

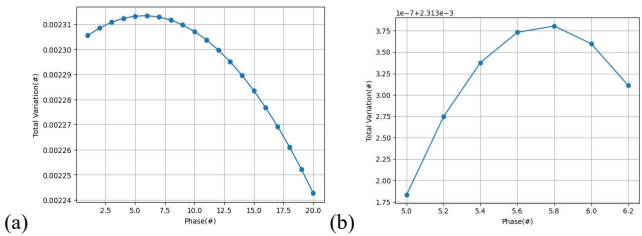

Fig 9. Total variation of I′ for different phase value (a) Phase range from 1 to 20, (b) Phase range from 5 to 6.

Figure 10 illustrates the results after adjusting the phase to minimize sharp features in the Q′ image, which has been converted from voltage to Tesla. The phase value (ϕ) is set at 5.8$^0$, as indicated in Eq. (2). This outcome can be quantified using field range enhancement (I-channel) and field range reduction (Q-channel), similar to the search efficiency and completeness equations, respectively mentioned in [13]. The adjusted phase, considering maximum and minimum field values for I′ and Q′ images, shows an improvement in sharpness of 0.3% and a reduction in sharpness by 25%, respectively. The resulting I′ image (Figure 10(a)) is sharper than the Q′ image (Figure 10(b)), as expected.

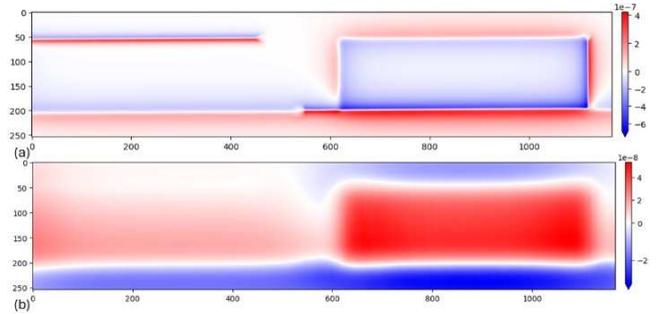

Fig 10. Adjusted phase (a) I′ image, (b) Q′ image

The raster scanned image exhibits a rotation of 0.3$^0$, which may result from skewed x and y imaging axes, or the x and y wires not being orthogonal, as illustrated in Figure 11(a). Figure 11(b) demonstrates the image transformation (T) of the I′ image, achieved by applying a 0.3$^0$ rotation. Figure 11(c) reveals the difference between Figures 11(a) and 11(b), highlighting features that should not have been transformed but were. In contrast, an affine transformation incorporating a 0.3$^0$ skew in the vertical direction (as shown in Figure 11(d)), followed by calculating the image difference with the original I′ image (Figure 11(a)), results in better alignment along the y-direction, as depicted in Figure 11(e).

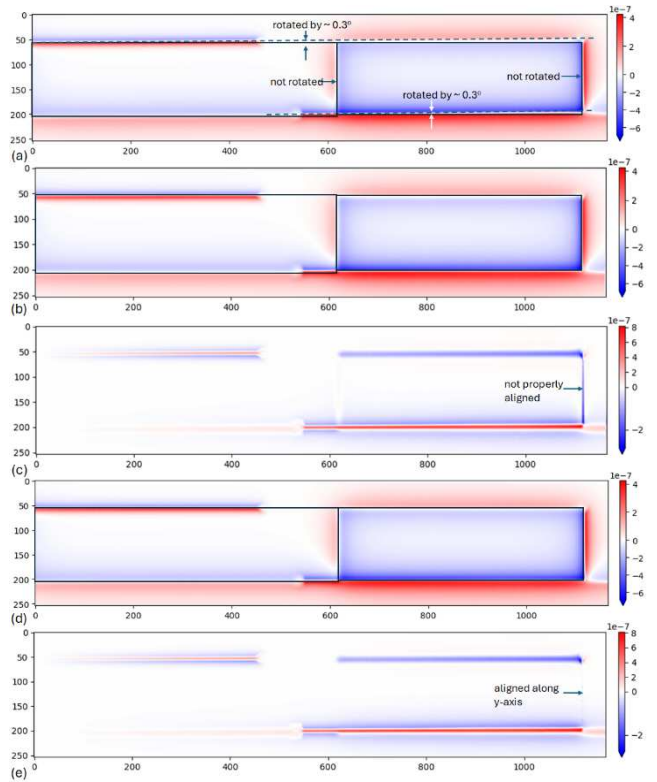

Fig. 11. (a) Original I′ image, (b) Aligned image with rotation of 0.3$^0$, (c) Image difference of (a) and (b), (d) Aligned image with skew of 0.3$^0$, (e) Image difference of (a) and (d)

An FFT-based inversion technique with a hard cut-off filter is used to convert field data into a two-dimensional current density distribution. Figure 12 illustrates the conversion of the aligned I' magnetic field image into a magnetic current image. This includes three representations: the current direction along the x-axis (Figure 12(a)), the current direction along the y-axis (Figure 12(b)), and the overall current density (Figure 12(c)). These images help us identify the current paths. When a magnetic field appears at the edge of the image, it can introduce artefacts, which can be observed on the right side of Figures 12(b) and 12(c).

In the FFT method, two important tunable parameters, $z$ and $k_w$, significantly influence the resulting current distribution. Figure 13 shows the effects of varying the cut-off ($k_w$) values at a fixed $z$ value of 120 μm (as indicated in Table I). The cut-off values varied from $1/z$ to $5/z$, representing Figures 13(a) through 13(e). Among these, the cut-off value of $3/z$ (as shown in Figures 12 and 13(b)) proves to be the most effective in accurately locating the current path compared to the other cut-off values.

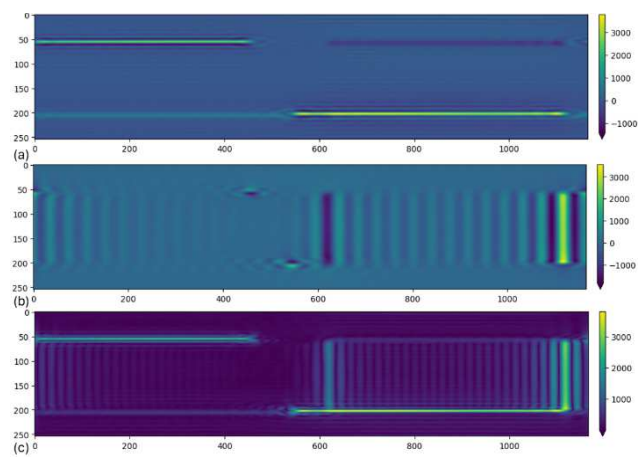

Fig. 12. Magnetic current image, (a) current flow direction along x-axis, (b) current flow direction along y-axis, (c) current density.

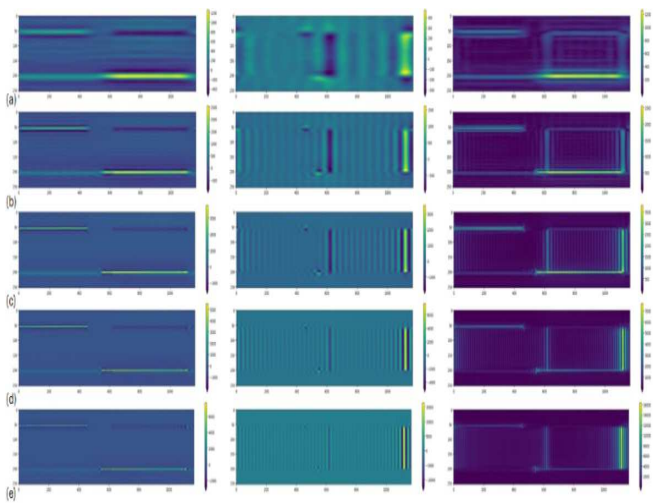

Fig. 13. Analysis of cut-off ($k_w$) values, (a) cut-off at $k_w=1/z$, (b) cut-off at $k_w=2/z$, (c) cut-off at $k_w=3/z$, (d) cut-off at $k_w=4/z$, (e) cut-off at $k_w=5/z$.

## IV. CONCLUSIONS

In this paper, we experimentally demonstrate the acquisition of magnetic field imaging using a SQUID microscopy scanner on a 3D spiral sample. We applied a novel approach called spatial-physics informed model (SPIM) to enhance the quality of the magnetic images. This approach addresses the effects of eddy currents, spatially aligns the images to correct for skew, and converts magnetic field data into magnetic current. The SPIM method shows several improvements: i) the sharpness of the processed I-channel improves by 0.3%, while the Q-channel experiences a sharpness reduction of 25%; ii) the image alignment through affine transformation, which addresses the skew effect of $0.3^0$, visually outperforms the adjustment for rotation; iii) applying an FFT with a cut-off value of 25 mm$^{-1}$ (where $k_w=3/z$ and $z=120$ μm) provides a feasible current density and direction. While this study primarily focuses on improving lateral resolution with a known $z$ value, there is potential for future extensions of the model to incorporate both lateral and vertical resolutions, as well as to investigate samples with defects for fault isolation [14].

ACKNOWLEDGMENT

This study is supported by the Machine Learning Guided Failure Analysis & Diagnostic Capability Development for Next Generation 3D-IC Packaging at A*STAR via the IAF-PP by the Agency for Science, Technology and Research under Grant No. M23K8a0050.